\documentclass[10pt,letterpaper,twocolumn]{article} 

\usepackage{ol2}
\usepackage[draft]{hyperref}
\usepackage{amsmath}

\begin{document}

\twocolumn[ 

\title{Synthetic gauge fields for light beams in optical resonators}


\author{Stefano Longhi}

\address{Dipartimento di Fisica, Politecnico di Milano and Istituto di Fotonica e Nanotecnologie del Consiglio Nazionale delle Ricerche, Piazza L. da Vinci 32, I-20133 Milano, Italy (stefano.longhi@polimi.it)}

\begin{abstract}
A method to realize artificial magnetic fields for light waves trapped in passive optical cavities with anamorphic optical elements is theoretically proposed. In particular, when a homogeneous magnetic field is realized, a highly-degenerate Landau level structure for the frequency spectrum of the transverse resonator modes is obtained, corresponding to a cyclotron motion of the optical cavity field. This can be probed by transient excitation of the passive optical resonator.   
\end{abstract}

\ocis{140.4780, 070.5753, 000.1600}


 ] 

\noindent 
{\it Introduction.} Realizing magnetic effects with photons at small spatial scales is a challenge. However, it would be extremely useful from both fundamental and applied aspects, enabling for instance the possibility to study quantum Hall effects for light, to built compact integrated optical isolators and to realize robust light transport via topologically-protected states \cite{M0,M1,M2,M3,M4,M5}. One possibility relies on the of synthesis of artificial gauge fields, which mimic magnetic or pseudo-magnetic fields for photons. This has been suggested and experimentally demonstrated in a series of recent works (see \cite{M5,M6,M7,M8,M9,M10,M11,M12,M13,M14,M15,M16,M17,M18} and references therein) using twisted \cite{M6,M7} or strained \cite{M8,M9} waveguide lattices, coupled-resonator optical waveguide structures \cite{M5,M10,M11,M12}, photonic lattices with suitable temporal modulation of the lattice parameters \cite{M13,M14,M15,M16}, and driven dissipative lattices of polaritons \cite{M17,M18}. Such previous studies have been mainly concerned with tight-binding lattice structures in either square or honeycomb geometries, where artificial gauge fields are generally synthesized by introduction of effective Peierls phases in the hopping rates among adjacent lattice sites. Another class of optical systems where light waves can behave like quantum particles in external potentials is provided by optical resonators \cite{M19}, where the transverse mode spectrum and corresponding beam dynamics can be controlled by proper design of the optical potential  \cite{M20,M21,M22,M23,M24}. 
\par  In this Letter we suggest a method to realize artificial magnetic fields for light waves trapped in passive optical cavities with anamorphic optical elements. In particular, for a homogeneous magnetic force a highly-degenerate Landau level structure for the frequency spectrum of the transverse resonator modes is obtained.\par
{\it Quantum particle in a magnetic field and Landau levels.}
The two-dimensional motion of a quantum particle of mass $m$ and charge $q$, constrained on the plane $(x,y)$ in the presence of external magnetic and electric fields $\mathbf{B}= \nabla \times \mathbf{A}$, $\mathbf{E}=-\nabla V$ described by the vectorial and scalar potentials $\mathbf{A}=A_x(x,y) \mathbf{u}_x+A_y(x,y) \mathbf{u}_y$   and $V(x,y)$, is governed by the Schr\"{o}dinger equation
\begin{equation}
i \hbar \frac{\partial \psi}{\partial t}=\frac{1}{2m} \left[ (\hat{p}_x-qA_x)^2+ (\hat{p}_y-qA_y )^2 \right] \psi +qV \psi \equiv \hat{\mathcal{H}} \psi
\end{equation}
for the particle wave function $\psi(x,y,t)$, where  $\hat{p}_x=-i \hbar \partial_x$ and $\hat{p}_y=-i \hbar \partial_y$ are the canonical momenta. In particular, for a uniform magnetic field $(\mathbf{B}=B_0 \mathbf{u}_z$, $\mathbf{E}=0$) in the Landau gauge one can assume $A_x=-B_0y$, $A_y=0$, $V=0$, leading to the Hamiltonian
$\hat{\mathcal{H}}=(1/2m) \left[ (\hat{p}_x-qA_x)^2+\hat{p}_y^2 \right]$.
As is well-known, the eigenfunctions of  such an Hamiltonian are a product of momentum eigenstates in the $x$ direction and harmonic oscillator eigenstates in the $y$ direction, namely  
\begin{eqnarray}
\psi  & = & \psi_n= \exp(ik_x x) H_n\left( \sqrt{\frac{m \omega_c}{\hbar}} (y-y_0) \right) \times \nonumber \\
& \times & \exp \left[- m \omega_c (y-y_0)^2 / 2 \hbar \right],
\end{eqnarray}
 where $\omega_c= qB_0/m$
 is the cyclotron frequency of classical motion and $y_0=\hbar k_x / (m \omega_c)$. The corresponding energies $E_n$, $\hat{\mathcal{H}} \psi_n=E_n \psi_n$, form a set of Landau levels, $E_n= \hbar \omega_c (n+1/2)$, which do not depend on the quantum momentum $k_x$ and are thus highly degenerate levels.  Here we show that an optical Schr\"{o}dinger equation like Eq.(1) arises in astigmatic optical cavities. \par
 \begin{figure}[htb]
\centerline{\includegraphics[width=8.4cm]{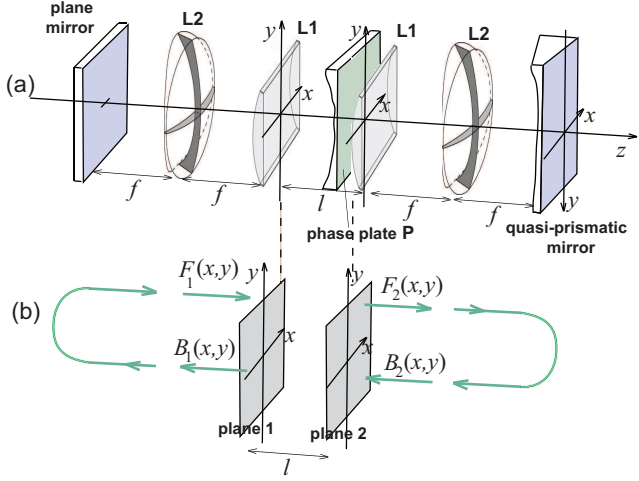}} \caption{ \small
(Color online)  (a) Schematic of the optical resonator with astigmatic optical elements that realizes a synthetic magnetic field in the transverse plane (L1: cylindrical lens; L2: spherical-cylindrical lens). (b) Round-trip propagation of the intracavity counter-propagating envelopes $F$ and $B$.}
\end{figure}

{\it Synthetic magnetic field in an optical resonator.} Let us consider the Fabry-Perot optical resonator shown in Fig.1(a). It comprises two high-reflectivity mirrors, a plane mirror (at the left side) and an aspherical (quasi-prismatic) mirror at the right side.  The aspherical mirror has a reflection coefficient $r(x,y)=\exp[i k x G(y)]$, where the function $G(y)$ is left arbitrary at this stage. Note that this is a quasi-prismatic  mirror, i.e. it has a strictly prismatic shape in the $x$ direction and an arbitrary structure in the $y$ direction.  Four lenses are placed between the two mirrors, 
namely two cylindrical lenses (L1) with focal length $f$ and two spherical-cylindrical lenses (L2) with focal lengths $f_x=f$ and $f_y=f/2$ in the $x$ and $y$ directions, respectively. The distances between the various adjacent lenses and mirrors are indicated in the figure. A thin plate P, acting as a {\it phase} filter, is also placed close to one of the two lenses L1, as depicted in Fig.1(a), with a transmission function $t(x,y)= \exp[i \varphi(x,y)]$. A monochromatic optical field $E(x,y,z,t)$ at frequency $\omega=k c=2 \pi c / \lambda$ trapped in the resonator is described by the superposition of  counter-propagating waves as $E=F(x,y,z) \exp(i \omega t-ikz)+B(x,y,z) \exp(i \omega t +i kz)$. The propagation of the  envelopes $F$ and $B$ at different planes inside the cavity can be readily obtained by application of the generalized Huygens-Fresnel integral for astigmatic beams \cite{M19,M25}. In particular, each of the two  spherical-cylindrical lenses L2 realize a one-dimensional Fourier transformation in the $x$ direction and an imaging step in the $y$ direction between the two planes far apart by $f$ from each lens. Using such a property, it can be readily shown that the following relations hold between the forward and backward propagating envelopes $F_{1,2}$ and $B_{1,2}$ at the two planes 1 and 2 shown in Fig.1(b)
\begin{equation}
B_2(x,y)=\hat{\mathcal{P}}_x F_2(x+fG(y),y) \; , \;\; F_1(x,y)=\hat{\mathcal{P}}_x B_1(x,y).
\end{equation}
where the symmetry operator $\hat{\mathcal{P}}_{x}$ is defined by $\hat{\mathcal{P}}_x \psi (x,y)=\psi(-x,y)$.
In writing Eq.(5), we assumed 100 \% mirror reflectance and no injected signal into the resonator; the effects of output coupling and signal injection will be discussed below. Taking into account diffraction and phase effects between the two planes 1 and 2, one can also write
\begin{equation}
F_2(x,y)  =  t(x,y) \hat{\mathcal{D}}F_1(x,y) \; , \; \;
B_1(x,y)  =  \hat{\mathcal{D}}  t(x,y)  B_2(x,y) 
 \end{equation}
where $ \hat{\mathcal{D}}=\exp[-il/ (2 k) \nabla^2_t]$ is the diffraction operator for a propagation length $l$, and $\nabla^2_t= \partial^2_x+\partial^2_y$ is the transverse Laplacian. From Eqs.(3) and (4) the resonator round-trip propagator $\hat{\mathcal{K}}$ at the plane 2, which connects the amplitude $F_2(x,y)$ of forward waves  at successive transits in the cavity can be readily derived and reads
\begin{equation}
\hat{\mathcal{K}}=t \hat{\mathcal{D}} \hat{\mathcal{P}_x} \hat{\mathcal{D}} t \hat{\mathcal{P}_x} \hat{\mathcal{S}} 
\end{equation}
where we have further introduced a shift operator $\hat{\mathcal{S}}$, defined by $\hat{\mathcal{S}} \psi (x,y)=\psi (x+fG(y),y)$.  Provided that a single longitudinal frequency of the cavity is excited and assuming the reference wave number $k$ to coincide with a longitudinal cavity wave number, the beam dynamics at successive transits inside the resonator can be described by the following map  
\begin{equation}
\phi^{(n+1)}(x,y)=\hat{\mathcal{K}}\phi^{(n)}(x,y)
\end{equation}
where $\phi^{(n)}(x,y) \equiv F_2^{(n)}(x,y)$ is the envelope of the progressive wave at plane $2$ and at the $n$-th transit in the cavity.
 \begin{figure}[htb]
\centerline{\includegraphics[width=8.6cm]{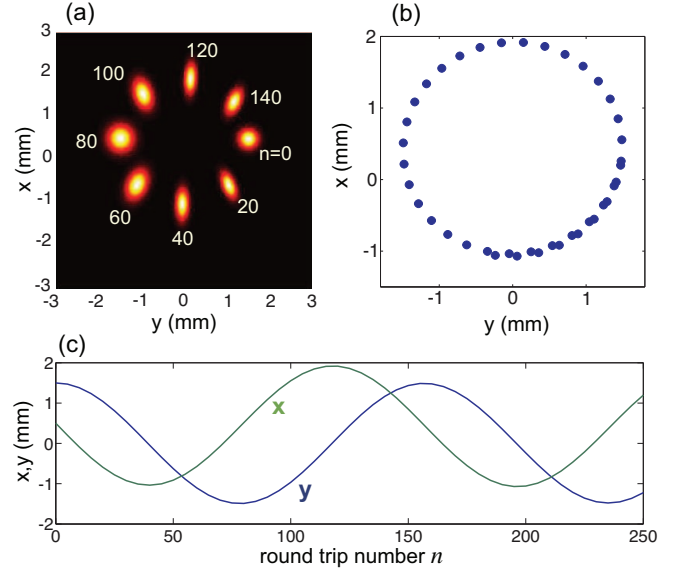}} \caption{ \small
(Color online)  Cyclotron motion of a Gaussian beam in the resonator under a homogeneous magnetic field: (a) Snapshots of the transverse intensity distribution at plane 2 at a few successive round trip numbers $n$. (b) and (c): trajectory of the beam center of mass $(x,y)$ in the transverse plane. Parameter values are given in the text.}
\end{figure}
 \begin{figure}[htb]
\centerline{\includegraphics[width=8.4cm]{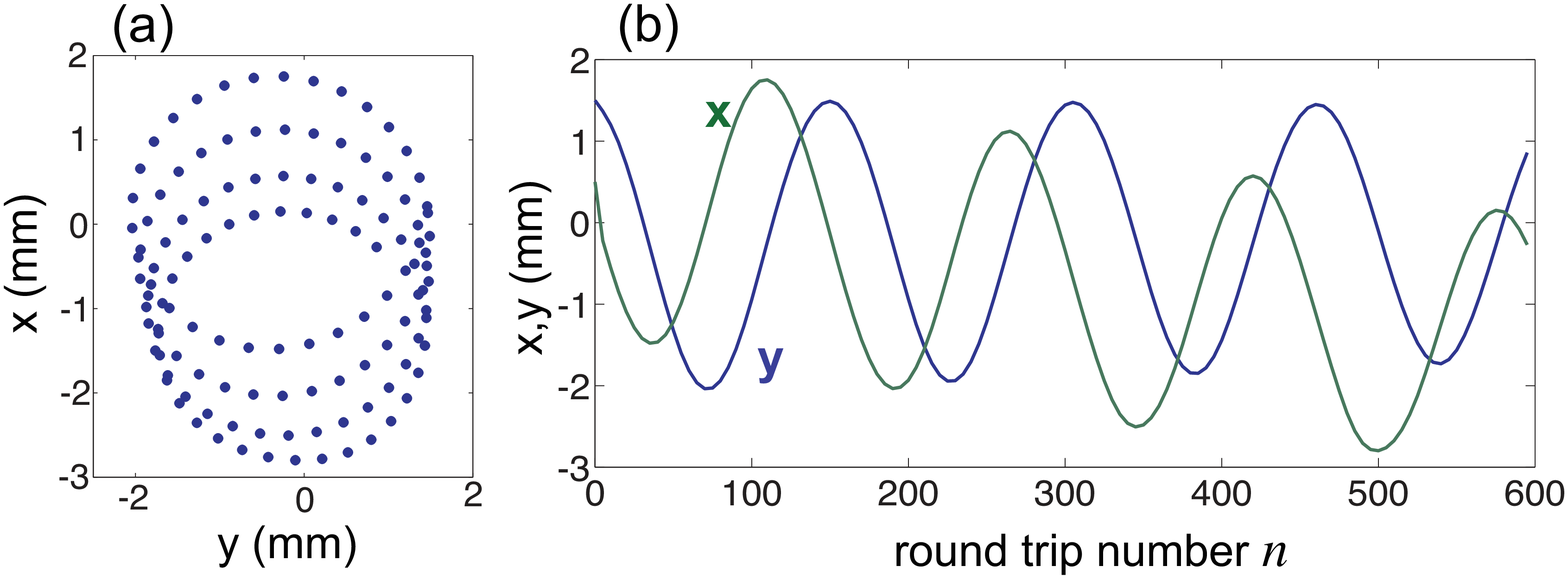}} \caption{ \small
(Color online)  Cyclotron motion of a Gaussian beam  under a gradient magnetic field. (a) and (b) show the trajectory of the beam center of mass $(x,y)$ in the transverse plane. Parameter values are given in the text.}
\end{figure}
 In the limit $ l \rightarrow 0$, $\varphi \rightarrow 0$, and $G \rightarrow 0$, i.e provided that the diffractive and the phase effects (introduced by the free-space propagation for a length $l$ and by the phase plate and prismatic mirror) are weak,
the resonator round-trip operator $\hat{\mathcal{K}}$  reduces at leading order to the identity operator $\hat{\mathcal{I}}$, because $\hat{\mathcal{P}_x} \hat{\mathcal{P}_x}=\hat{\mathcal{I}}$. 
In such a case, the field envelope $\phi^{(n)}(x,y)$ undergoes a slow change after each round trip and one can set $\hat{\mathcal{K}} \simeq \hat{\mathcal{I}}+ i \hat{\Omega}$,  where the leading order correction $i \hat{\Omega}$ to the identity operator is obtained from Eq.(5) by first-order power expansion of the operators $\hat{\mathcal{D}}$, $t$ and $\hat{\mathcal{S}}$. After continuation of the round-trip index $n$ and setting $\phi^{(n+1)}(x,y)- \phi^{(n)}(x,y) \simeq (\partial \phi / \partial n)$ with $\phi(x,y,n)=\phi^{(n)}(x,y)$, one then obtains the evolution equation (mean-field equation) $ i (\partial \phi/ \partial n)=- \hat{\Omega} \phi$, which can be cast in the following form
\begin{equation}
i \hbar \frac{\partial \psi} {\partial n}= \frac{1}{2m} \left( -i \hbar \frac{\partial}{\partial x} -qA_x \right)^2 \psi -\frac{\hbar^2}{2m} \frac{\partial^2 \psi}{\partial y^2}+ qV \psi.
\end{equation}
In Eq.(7) we have set $ \psi(x,y,n)= \phi^*(x,y,n)$ and
\begin{eqnarray}
\hbar  & = & 1/k = \lambda / (2 \pi)  \\
m & = & 1/(2l) \\
 qA_x  & = &  \frac{ f  G(y)}{2l} \\
  qV(x,y)  & = & - \frac{1}{k} [\varphi(x,y)+\varphi(-x,y)]-	\frac{f^2 G^2(y)}{4l}. \;\;\;\;\;\;\;
  \end{eqnarray}
  The mean-field equation (7) clearly corresponds to the Schr\"{o}dinger equation (1) for a quantum particle in an external electric and magnetic field with $A_y=0$, provided that the formal substitutions (8-11) are made in the Schr\"{o}dinger equation. Note that the round-trip number $n$ in Eq.(7) corresponds to the time variable normalized to the cavity round-trip time $T_R=2L_e/c$, where $L_e$ is the optical resonator length. Note also that the magnetic field is controlled by the phase profile $G(y)$ of the quasi-prismatic mirror, and turns out to be generally inhomogeneous in the $y$ direction. The scalar potential $V$ is in turn controlled by both the phase profile $G(y)$ and the phase profile $\varphi(x,y)$ of the phase plate P according to Eq.(11). In particular, with the choice
  \begin{equation}
  \varphi(x,y)=- k f^2 G^2(y)/(8l)
  \end{equation} 
the potential $V$ vanishes, and Eq.(7) describes the motion of a quantum particle in an inhomogeneous magnetic field. Note that $\varphi$ defined by Eq.(12) does not depend on $x$ and it thus describes the effect of a (generally aspherical) cylindrical lens. 
\par
{\it Transverse mode spectrum and beam dynamics.} 
 Let us first consider the case of a homogeneous magnetic field. This is obtained by assuming a linear gradient for $G$, i.e. $G(y)= \alpha y$,
which corresponds to a magnetic field $qB_0= \alpha f /(2l)$. Note that in this case a quadratic phase $\varphi$ [Eq.(12)] is required to cancel the scalar potential $V$, i.e. the phase plate P in Fig.1(a) can be simply replaced by a converging cylindrical lens with focal length $f'$ in the $y$ direction given by $f'= 2 l / (f \alpha)^2$. The transverse mode spectrum corresponds to a set of highly-degenerate Landau levels at frequencies $\omega_n=\omega+2 \pi \Delta \nu (1/2+n) $ with frequency separation
\begin{equation}
\Delta \nu=\frac{\omega_c}{2 \pi T_R}=\frac{\alpha f}{2 \pi T_R}= \frac{\alpha f} {2 \pi} \Delta \nu_{ax}
\end{equation}
where $n=0,1,2,...$ is the Landau level index and $\Delta \nu_{ax}=1/T_R=c/(2L_e)$ is the frequency separation of the cavity axial modes. The corresponding transverse modes are given by Eq.(2). Note that the Landau degeneracy of transverse modes is rather distinct than degeneracy found in the spectrum of other broad area optical cavities (such as plane-plane or confocal resonators) generally employed to study transverse patterns in lasers and nonlinear optical systems \cite{M26,M27}. A distinctive feature here is the possibility to observe a cyclotron motion as a result of mode degeneracy. This is shown, as an example, in the numerical simulations of Fig.2 for resonator parameters $f=4$ cm, $l=0.4$ cm, corresponding to a total cavity length $L_e \simeq  4f+l \simeq 16.4$ cm, probed at the wavelength $\lambda=633$ nm. Assuming $G(y)= \alpha y$ with $\alpha=1 \; {\rm m}^{-1}$, the frequency separation of the set of degenerate Landau modes is given by $\Delta \nu \simeq 5.82$ MHz, corresponding to a cyclotron motion that occurs after $n_c = 2 \pi / (\alpha f)  \simeq 157$ round trips in the resonator, i.e. with a period $\tau_c=n_cT_R= 1 / \Delta \nu \simeq 172$ ns.  Figure 2(a) shows snapshots of the transverse beam intensity evolution at successive round trips as obtained by numerical solution of the map (6), taking as an initial condition an off-axis Gaussian field distribution with spot size $w$ and transverse displacements $x_1$ and $y_1$, namely $\psi^{(0)}  \propto \exp[-(x-x_1)^2/w^2-(y-y_1)^2/w^2]$ with $w= 300\; \mu$m, $x_1=0.5$ mm and $y_1=1.5$ mm. The trajectory of the beam center of mass in the transverse $(x,y)$ plane, taken every 5 round trips, is depicted in Fig.2(b) and clearly corresponds to a nearly circular path of radius $R \simeq y_1$,  according to a semiclassical analysis of the Hamiltonian $\hat{\mathcal{H}}$ with the chosen Landau gauge. Interestingly, for a weak inhomogeneous magnetic field in the $y$ direction, corresponding to $(d^2G/dy^2) \neq 0$,  the beam path is given by the superposition of a relatively fast circular motion around a point (the guiding center) and a relatively slow drift of this point along the $x$-axis. The weak drift of the guiding center is shown in Fig.3 for a quadratic phase profile $G(y)=\alpha y- \beta y^2$ with $\alpha=1 \; {\rm m}^{-1}$ and $\beta=30 \; {\rm m}^{-2}$; other parameters are as in Fig.2. \par The cyclotron motion depicted in Figs.2 and 3 could be experimentally observed by monitoring the intensity distribution in a resonator driven by an external beam. Indicating by $T \ll 1$ the transmittance of the coupling plane mirror [at the left side in Fig.1(a)] and by $E_{in}^{(n)}(x,y)$ the injected field amplitude at a frequency in resonance with one axial mode of the resonator and slowly varying over one round trip, the evolution of the intracavity field is now described by the following modified map [compare with Eq.(6)]
\begin{equation}
\phi^{(n+1)}(t)=\hat{\mathcal{K}} \phi^{(n)}+ \sqrt{T} E_{in}^{(n)}-\frac{T}{2} \phi^{(n)}.
\end{equation}
Similarly, the mean-field equation (7) takes the form
\begin{eqnarray}
i \hbar \frac{\partial \psi} {\partial n} & = & \frac{1}{2m} \left( -i \hbar \frac{\partial}{\partial x} -qA_x \right)^2 \psi -\frac{\hbar^2}{2m} \frac{\partial^2 \psi}{\partial y^2}+ V \psi  \nonumber \\
& - & i \hbar \frac{T}{2} \psi+ i \hbar \sqrt{T}  E_{in}.
\end{eqnarray}
As an example, Fig.4 shows the evolution of the beam position inside the resonator when a Gaussian beam is coupled into the cavity via the plane mirror  ($T=1 \%$ transmittance) and slowly switched off. In the simulations, we assumed  $E_{in}^{(n)}=f(n) \exp[-(x-x_1)^2/w^2-(y-y_1)^2/w^2]$, where the function $f(n)=(1/2)[1-{\rm tanh}((n-n_0)/ \Delta n)]$ describes the switch off of the the injected signal at the round trip $n_0$ [see the dashed curve in Fig.4(b)]. Parameter values used in the simulations are the same as those in Fig.2, except for $\alpha= 2 \; {\rm m}^{-1}$. For $n<n_0$, the intracavity field is stationary and it shows a characteristic annular shape [Fig.4(a)], which is the signature of a cyclotron dynamics: the annular shape at steady-state operation arises from a balance between circular drift induced by the synthetic gauge field and continuous signal feeding at $x=x_1$, $y=y_1$. As the external field is switched off ($n>n_0$) and signal feeding vanishes, the intracavity field decays in time undergoing a cyclotron motion, see Fig.4(b). Note that the cyclotron period is  $1 / \Delta \nu \simeq 86$ ns, which is smaller than the cavity decay rate $T_R/T \simeq 109.4$ ns. Hence the spiraling of the decaying output light beam when the holding beam is switched off should be observable by time-resolved transverse pattern evolution measurements \cite{M28,M29}. 
 \begin{figure}[htb]
\centerline{\includegraphics[width=8.6cm]{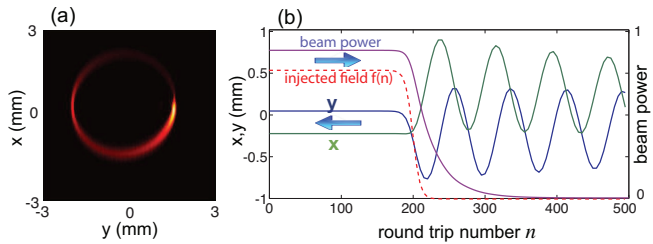}} \caption{ \small
(Color online)  Cyclotron motion in a resonator with injected Gaussian beam. (a) Stationary intensity distribution at plane 2 under signal injection ($n<n_0$). (b) Trajectory of the beam center of mass $(x,y)$ when the injected field is switched off. The behavior of the intracavity power (in arbitrary units) is also shown, together with the decaying amplitude  $f(n)=(1/2)[1-{\rm tanh}((n-n_0)/ \Delta n)]$ of the injected field (dashed curve) with $n_0=200$ and $\Delta n=10$. Other parameter values are given in the text.}
\end{figure} 
\par
{\it Conclusions.} Artificial magnetic fields for photons can be realized in the transverse plane of an optical resonators with astigmatic elements. Such a result  sheds new light into the rapidly growing field of synthetic magnetic fields in optical systems and might suggest further investigations. For example,
 the degenerate Landau level structure of the resonator can provide a platform for the study of laser mode competition and transverse pattern formation  \cite{M26,M27} in presence of a circular drift (cyclotron motion). Also, 
 by including a periodic phase mask in the cavity, the coherently-driven astigmatic resonator could implement
a driven-dissipative two-dimensional photonic lattice with a synthetic gauge field \cite{M18}, thus providing a viable route toward the experimental 
observation of optical analogues of the anomalous and the (integer) quantum Hall effect.

\newpage

\footnotesize {\bf References with full titles}\\
\\
1. L. Lu, J.D. Joannopoulos, and M. Soljacic, "Topological photonics", Nature Photon. {\bf 8}, 821 (2014).\\ 
2. S. Raghu and F.D.M. Haldane, "Analogs of quantum-Hall-effect edge states in photonic crystals", Phys. Rev. A {\bf 78}, 033834 (2008).\\
3. F.D.M. Haldane and S. Raghu, "Possible realization of directional optical waveguides in photonic crystals with broken time-reversal symmetry",
Phys. Rev. Lett. {\bf 100}, 013904 (2008).\\
4. Z. Wang, Y. Chong, J.D. Joannopoulos,  and M. Soljacic,  "Reflection-free one-way edge modes in a gyromagnetic photonic crystal", Phys. Rev. Lett.
{\bf 100}, 013905 (2008).\\
5. Z. Wang, Y. Chong, J.D. Joannopoulos, and M. Soljacic, "Observation of unidirectional backscattering-immune topological electromagnetic states",
Nature {\bf 461}, 772Ð775 (2009).\\
6.  M. Hafezi, E.A. Demler, M.D. Lukin, and J.M. Taylor, "Robust optical delay lines with topological protection", Nature Phys. {\bf 7}, 907 (2011).\\
7. S. Longhi, "Bloch dynamics of light waves in helical optical waveguide arrays", Phys. Rev. B {\bf 76}, 19511 (2007).\\
8. M.C. Rechtsman, J.M.  Zeuner, Y. Plotnik, Y.Lumer, D. Podolsky, F. Dreisow, S. Nolte, M. Segev, and A. Szameit, "Photonic Floquet Topological Insulators", Nature {\bf 496}, 196 (2013).\\
9. M.C. Rechtsman, J.M. Zeuner, A. T\"{u}nnermann, S. Nolte, M. Segev, and A. Szameit, "Strain-induced pseudomagnetic field and photonic Landau levels in dielectric structures", Nature Photon. {\bf 7}, 153 (2013).\\ 
10. H. Schomerus and N.Y. Halpern, "Parity Anomaly and Landau-Level Lasing in Strained Photonic Honeycomb Lattices", Phys. Rev. Lett. {\bf 110}, 013903 (2013).\\
11. M. Hafezi, S. Mittal, J. Fan, A. Migdall, and J.M. Taylor, "Imaging Topological Edge States in Silicon Photonics",  Nature Photon. {\bf 7}, 1001 (2013).\\
12. G. Q. Liang and Y. D. Chong, "Optical resonator analog of a two-dimensional topological insulator", Phys. Rev. Lett. {\bf 110}, 203904 (2013).\\
13. S. Mittal, J. Fan, S. Faez, A. Migdall, J.M. Taylor, and M. Hafezi, "Topologically Robust Transport of Photons in a Synthetic Gauge Field",  Phys. Rev. Lett. {\bf 113}, 087403 (2014).\\
14. K. Fang, Z. Yu, and S. Fan, "Realizing effective magnetic field for photons by controlling the phase of dynamic modulation", Nature Photon. {\bf 6}, 782 (2012).\\
15. S. Longhi, "Effective magnetic fields for photons in waveguide and coupled resonator lattices", Opt. Lett. {\bf 38}, 3570 (2013).\\
16. S. Longhi, "AharonovÐBohm photonic cages in waveguide and coupled resonator lattices by synthetic magnetic fields", Opt. Lett. {\bf 39}, 5892 (2014).\\
17. L.D. Tzuang,K. Fang, P. Nussenzveig, S. Fan, and M. Lipson, "Non-reciprocal phase shift induced by an effective magnetic flux for light", Nature Photon. {\bf 8}, 701 (2014).\\
18. O. Umucallar and I. Carusotto, "Artificial gauge field for photons in coupled cavity arrays", Phys. Rev. A {\bf 84}, 043804 (2011).\\
19. T. Ozawa and I. Carusotto, "Anomalous and Quantum Hall Effects in Lossy Photonic Lattices", Phys. Rev. Lett. {\bf 112}, 133902 (2014).\\
20. A.E. Siegman, {\it Lasers} (University Science Books, Mill Valley, CA, 1986).\\
21. C. Pare, L. Gagnon, and P. A. Belanger, "Aspherical laser resonators: An analogy with quantum mechanics", Phys. Rev. A {\bf 46}, 4150 (1992).\\
22. M. Kuznetsov, M. Stern, and J. Coppeta, "Single transverse mode optical resonators", Opt. Express {\bf 13}, 171 (2004).\\
23. B.T. Torosov, G. Della Valle, and S. Longhi, "Imaginary Kapitza pendulum", Phys. Rev. A {\bf 88}, 052106 (2013).\\
24. S. Ngcobo, I. Litvin, L. Burger, and A. Forbes, "A digital laser for on-demand laser modes", Nature Commun. {\bf 4}, 2289 (2013).\\
25. S. Longhi, "Fractional Schr\"{o}dinger equation in optics", Opt. Lett. {\bf 40}, 1117 (2015).\\ 
26. P. Baues, "The Connection of Geometrical Optics with the Propagation of Gaussian Beams and the Theory of Optical Resonators", Opto-Electronics {\bf 1}, 103 (1969).\\
27. K. Staliunas and V. Sanchez-Morcillo, {\it Transverse Patterns in Nonlinear Optical Resonators} (Springer, Berlin 2003).\\
28. L. Lugiato, F. Prati, and M. Brambilla, {\it Nonlinear Optical Systems} (Cambridge University Press, Cambridge, 2015).\\ 
29. G. Yao, S.H. Zhou, P. Wang, K.K. Lee, and Y.C. Chen, "Dynamics of transverse mode in self-Q-switched solid-state lasers", Opt. Commun. {\bf 114}, 101 (1995).\\
30. F. Encinas-Sanz, S. Melle, and O.G. Calderon, "Time-Resolved Dynamics of Two-Dimensional Transverse Patterns in Broad Area Lasers", Phys. Rev. Lett. {\bf 93}, 213904 (2004).\\


\begin{thebibliography}{99}




\bibitem{M0}
L. Lu, J.D. Joannopoulos, and M. Soljacic, Nature Photon. {\bf 8}, 821(2014).

\bibitem{M1}
S. Raghu and F.D.M. Haldane, Phys. Rev. A {\bf 78}, 033834 (2008).

\bibitem{M2}
F.D.M. Haldane and S. Raghu, Phys. Rev. Lett. {\bf 100}, 013904 (2008).

\bibitem{M3}
Z. Wang, Y. Chong, J.D. Joannopoulos,  and M. Soljacic, Phys. Rev. Lett.
{\bf 100}, 013905 (2008).

\bibitem{M4}
Z. Wang, Y. Chong, J.D. Joannopoulos, and M. Soljacic,  Nature {\bf 461}, 772 (2009).

\bibitem{M5}
M. Hafezi, E.A. Demler, M.D. Lukin, and J.M. Taylor, Nature Phys. {\bf 7}, 907 (2011).

\bibitem{M6}
S. Longhi,  Phys. Rev. B {\bf 76}, 19511 (2007).

\bibitem{M7}
M.C. Rechtsman, J.M.  Zeuner, Y. Plotnik, Y.Lumer, D. Podolsky, F. Dreisow, S. Nolte, M. Segev, and A. Szameit, Nature {\bf 496}, 196 (2013).

\bibitem{M8} 
M.C. Rechtsman, J.M. Zeuner, A. T\"{u}nnermann, S. Nolte, M. Segev, and A. Szameit,  Nature Photon. {\bf 7}, 153 (2013).

\bibitem{M9}
H. Schomerus and N.Y. Halpern, Phys. Rev. Lett. {\bf 110}, 013903 (2013).

\bibitem{M10}
M. Hafezi, S. Mittal, J. Fan, A. Migdall, and J.M. Taylor, Nature Photon. {\bf 7}, 1001 (2013).

\bibitem{M11}
G. Q. Liang and Y. D. Chong, Phys. Rev. Lett. {\bf 110}, 203904 (2013).

\bibitem{M12}
S. Mittal, J. Fan, S. Faez, A. Migdall, J.M. Taylor, and M. Hafezi,  Phys. Rev. Lett. {\bf 113}, 087403 (2014).

\bibitem{M13}
K. Fang, Z. Yu, and S. Fan, Nature Photon. {\bf 6}, 782 (2012).

\bibitem{M14}
S. Longhi, Opt. Lett. {\bf 38}, 3570 (2013).

\bibitem{M15}
S. Longhi, Opt. Lett. {\bf 39}, 5892 (2014).
	
\bibitem{M16}
 L.D. Tzuang,K. Fang, P. Nussenzveig, S. Fan, and M. Lipson, Nature Photon. {\bf 8}, 701 (2014).
 
 \bibitem{M17}
O. Umucallar and I. Carusotto, Phys. Rev. A {\bf 84}, 043804 (2011)

\bibitem{M18}
T. Ozawa and I. Carusotto, Phys. Rev. Lett. {\bf 112}, 133902 (2014).

\bibitem{M19}
A.E. Siegman, {\it Lasers} (University Science Books, Mill Valley, CA, 1986).

\bibitem{M20}
C. Pare, L. Gagnon, and P. A. Belanger, Phys. Rev. A {\bf 46}, 4150 (1992).

\bibitem{M21}
M. Kuznetsov, M. Stern, and J. Coppeta, Opt. Express {\bf 13}, 171 (2004).

\bibitem{M22}
B.T. Torosov, G. Della Valle, and S. Longhi, Phys. Rev. A {\bf 88}, 052106 (2013).

\bibitem{M23}
 S. Ngcobo, I. Litvin, L. Burger, and A. Forbes, Nature Commun. {\bf 4}, 2289 (2013).

\bibitem{M24}
S. Longhi, Opt. Lett. {\bf 40}, 1117 (2015).

\bibitem{M25}
 P. Baues, Opto-Electronics {\bf 1}, 103 (1969).

\bibitem{M26}
K. Staliunas and V. Sanchez-Morcillo, {\it Transverse Patterns in Nonlinear Optical Resonators} (Springer, Berlin 2003).

\bibitem{M27}
L. Lugiato, F. Prati, and M. Brambilla, {\it Nonlinear Optical Systems} (Cambridge University Press, Cambridge, 2015).

\bibitem{M28}
G. Yao, S.H. Zhou, P. Wang, K.K. Lee, and Y.C. Chen, Opt. Commun. {\bf 114}, 101 (1995).

\bibitem{M29}
F. Encinas-Sanz, S. Melle, and O.G. Calderon, Phys. Rev. Lett. {\bf 93}, 213904 (2004).









\end{thebibliography}
\end{document}